\documentclass[prl,aps,showpacs,floats,letterpaper,floatfix,
groupedaddress,superscriptaddress
,nofootinbib
,reprint
,twocolumn
]{revtex4}
\usepackage{amsmath}
\usepackage{amsfonts}
\usepackage{bm}
\usepackage{graphicx}

   \newcommand{\aj}{Astron. J.}

\def\v1v2{{\bf v}_1 \cdot {\bf v}_2}
\allowdisplaybreaks

\begin{document}

\title{A new general relativistic contribution to Mercury's perihelion advance}

\author{
Clifford M.~Will} \email{cmw@phys.ufl.edu}
\affiliation{Department of Physics, University of Florida, Gainesville, Florida 32611, USA}
\affiliation{GReCO, Institut d'Astrophysique de Paris, CNRS,\\ 
Universit\'e Pierre et Marie Curie, 98 bis Boulevard Arago, 75014 Paris, France}

\date{\today}

\begin{abstract}
We point out the existence of a new general relativistic contribution to the perihelion advance of Mercury that, while smaller than the contributions arising from the solar quadrupole moment and angular momentum, is 100 times larger than the second-post-Newtonian contribution.  It arises in part from  relativistic ``cross-terms'' in the post-Newtonian equations of motion between Mercury's interaction with the Sun and with the other planets, and in part from an interaction between Mercury's motion and the  gravitomagnetic field of the moving planets.   At a few parts in $10^6$ of the leading general relativistic precession of 42.98 arcseconds per century, these effects are likely to be detectable by the BepiColombo mission to place and track two orbiters around Mercury, scheduled for launch around 2018.
\end{abstract}

\maketitle

\section{Introduction}
\label{sec:intro}

The perihelion advance of Mercury is one of the iconic tests of Einstein's general theory of relativity.  The story began as a crisis of 19th-century Newtonian dynamics, when Le Verrier pointed out in 1859 \cite{1859AnPar...5....1L} that a tally of the perturbations of Mercury's orbit induced by the Newtonian gravitational attraction of the other planets fell short of  accounting for the observed advance of the perihelion by an amount that was 43 arcseconds per century (here we use the modern value).
Notwithstanding imaginative attempts to resolve this discrepancy by postulating an intramercurial planet (dubbed ``Vulcan'') or by tweaking the Newtonian inverse square law, it was still an unsolved problem when Einstein began his quest for relativistic theory of gravity in 1907 \cite{1982mpfl.book.....R}.   In fact, he used the perihelion advance problem as a filter for his various preliminary theories.    Although he was already becoming dissatisfied with the theoretical properties of the {\em entwurf}, or ``draft'' theory that he had developed in 1912 with Marcel Grossmann, he finally rejected it because it failed to give the right answer for the Mercury discrepancy.    In November 1915, when everything seemed to be falling into place theoretically for his latest attempt, the tipping point occurred when he saw that the theory gave the correct value for the missing perihelion advance.   He later wrote that this discovery gave him ``palpitations of the heart''  (see, for example Sec.\ 14c of \cite{1982sils.book.....P}).

The perihelion advance became a ``hot topic'' again in the 1960s, when Dicke and collaborators claimed to have shown, through observations of the shape of the solar disk, that the Sun was sufficiently oblate that the Newtonian contributions of the modified solar potential would contribute four arcseconds per century (as/cy) to the perihelion advance, thus invalidating general relativity, and supporting Brans and Dicke's recent scalar-tensor theory of gravity, which predicted only about 39 as/cy \cite{1967PhRvL..18..313D}.

Since the 1970s the perihelion advance has entered the pantheon of high-precision confirmations of general relativity.  This was made possible by developments on many fronts: high-precision radar tracking of planets and spacecraft; improvements in our knowledge of planetary and asteroid masses; precise measurements of the Earth-Moon orbit using lunar laser ranging; development of computer codes for obtaining precise orbits of the planets, major asteroids and spacecraft; and helioseismology, which yielded credible values for the solar quadrupole moment.   

Another development was the adoption of the post-Newtonian limit of general relativity and later of the parametrized post-Newtonian (PPN) formalism 
\cite{1968PhRv..169.1017N,1971ApJ...163..611W,tegp,tegp2}
as the foundation for analyzing solar system data
\cite{1971mfdo.book.....M,moyer2000}.  The PPN formalism provides, among other tools, a set of $N$-body equations of motion, valid to the first post-Newtonian (PN) order [$O(v/c)^2$ beyond Newtonian theory], expressed in terms of a set of dimensionless parameters, $\gamma,\, \beta, \, \dots$ whose values depend on the theory of gravity being used.  This made it possible to analyze all available solar system data, both historical and current, in a uniform manner.  Through such estimation procedures as least squares, one could obtain estimates for the values and uncertainties in the parameters of the problem, such as masses and orbit elements of planets, the quadrupole moment of the Sun, and relativistic parameters such as $\gamma$ and $\beta$, and to understand correlations among them.   The main improvements to the estimates of $\gamma$ came from data in which the tracking signal passes close to the Sun, thus experiencing the Shapiro time delay, which depends on $(1+\gamma)/2$.  Such was the case for analyses that included data from the 2003 cruise phase of the Cassini mission to Saturn, which yielded $\gamma - 1= (2.1 \pm 2.3) \times 10^{-5}$ \cite{2003Natur.425..374B}.  The parameter $\beta$ is sensitive to the perihelion advance, which depends on $(2 + 2\gamma - \beta)/3$, and to the ``Nordtvedt'' effect in lunar laser ranging, which depends primarily on $4\beta - \gamma -3$.  Recall that $\gamma = \beta =1$ in general relativity (GR).  

A major advance in measuring the parameters of the perihelion advance was made by exploiting Mercury MESSENGER.   
In 2011, MESSENGER became the first spacecraft to orbit Mercury, and range and Doppler measurements of the orbiter were made until the spacecraft ended its mission in 2015 with a controlled crash on the surface of Mercury.   By 2013, MESSENGER data had already led to dramatically improved knowledge of Mercury's orbit.   Analyses of all the available data yielded  bounds on $\gamma$ and $\beta$ given by $\gamma-1 = (-0.3 \pm 2.5) \times 10^{-5}$ and $\beta -1 = (0.2 \pm 2.5) \times 10^{-5}$ \cite{2011CeMDA.111..363F,2014A&A...561A.115V,2015CeMDA.123..325F}.    The analyses also yielded an estimate for the solar quadrupole moment $J_2 = (2.4 \pm 0.2) \times 10^{-7}$, consistent with the results from helioseismology.    More recent analyses yielded comparable results \cite{2017AJ....153..121P,2018NatCo...9..289G}.

Improved measurements of $\beta$, down to the level of parts per million \cite{2002PhRvD..66h2001M,2007PhRvD..75b2001A}, may be possible using data from  the joint European-Japanese BepiColombo project to place two orbiters around Mercury
 \cite{2010P&SS...58....2B}, scheduled for launch in late 2018.
The purpose of this Letter is to point out that, at the level of parts per million, there is a new general relativistic effect on Mercury's perihelion that has not been calculated explicitly heretofore, although it is implicit in the PPN $N$-body equations of motion mentioned above.  This is the effect of post-Newtonian ``cross-terms'' in the equations of motion \cite{2014PhRvD..89d4043W}.

To understand PN cross terms, consider a hierarchical triple system, consisting of an inner binary system (the Sun and Mercury in this case) and a distant third body.  
At Newtonian order, the relative acceleration between Mercury and the Sun has terms of order
$Gm/r^2$ and $Gm_3 r/R^3$, where $m$ and $r$ are the mass and separation of the Sun-Mercury system, and where we have expanded the effect of the  external body at distance $R$ and with mass $m_3$ to only quadrupole order.   In the post-Newtonian approximation, each of these terms
comes from a potential $Gm/r$ and $Gm_3 r^2/R^3$, which then leads to a dimensionless relativistic correction factor $Gm/rc^2$ and $Gm_3 r^2/R^3c^2$.  Here $G$ and $c$ are the gravitational constant and the speed of light, respectively.
Thus, in addition to the two Newtonian acceleration terms and the usual PN corrections to the acceleration within the Sun-Mercury system, of order $G^2m^2/r^3 c^2$, we also include ``cross terms'' between PN and third-body effects, of the form $(Gm/r^2)\times(Gm_3r^2/R^3c^2)$ or $(Gm_3r/R^3)\times(Gm/rc^2)$, both of which scale as $G^2 mm_3/R^3c^2$.  Thus we are including the relativistic effect of the third body's potential on the Newtonian acceleration due to the Sun, and the relativistic effect of the Sun's potential on the perturbing acceleration due to the third body.  Relative to the dominant Newtonian acceleration by the Sun, these cross terms have a dimensionless scale given by $Gm_3 r^2/R^3 c^2 \sim [(m_3/m)(r/R)^3 ] [Gm/rc^2]$.  Thus we would expect on dimensional grounds that, if these cross-term perturbations induce a perihelion advance for Mercury, it would be of order $Gm/rc^2 \sim 10^{-7}$ times the Newtonian advance induced by the other planets.  But this advance is $\sim 530$ as/cy, an order of magnitude larger than the standard relativistic advance.  Thus we might expect the contribution of PN cross terms to be at the level of parts per million of the GR effect, exactly the regime that will be explored by BepiColombo.  

Another PN cross-term that turns out to be relevant is an interaction between the velocity $\bm{v}$ of Mercury  and the ``gravitomagnetic (GM) field'' generated by the ``mass current'' of the moving third body.  This interaction is proportional to $Gm_3 {V}_3 {v}/c^2 R^2$.  With $v \sim (Gm/r)^{1/2}$ and $V_3 \sim (Gm/R)^{1/2}$, this scales  as $[(m_3/m)(r/R)^{5/2} ] [Gm/rc^2]$ relative to the Newtonian solar acceleration, and could lead to a contribution to the perihelion advance of comparable size. 

A detailed calculation, to be described in the next section, confirms these expectations.  For a third body in a circular orbit that is coplanar with the Sun-Mercury system, the advance per orbit of the perihelion of Mercury is given by
\begin{align}
\Delta \varpi &= \frac{6\pi Gm}{c^2 p} + \frac{3\pi}{2} \frac{m_3}{m} \left ( \frac{a}{R} \right )^3 (1-e^2)^{1/2} 
\nonumber \\
& \quad
+ \frac{3\pi}{4} \frac{Gm_3 a^2}{c^2 R^3} \frac{28 + 47e^2}{(1-e^2)^{3/2}} + 4\pi  \frac{Gm_3}{c^2 a} \left ( \frac{a}{R} \right )^{5/2} \,, 
\label{eq1:delom1}
\end{align}
where $\varpi$ is the perihelion measured from a fixed reference direction, $a$ and $e$ are the semimajor axis and eccentricity of Mercury's orbit and $p=a(1-e^2)$.   The first term is the standard general relativistic precession, the second is the Newtonian precession induced by the third body, and the third is the cross-term effect arising from the coupling between the solar and third-body potentials.  The final term in Eq.\ (\ref{eq1:delom1}) comes from the gravitomagnetic cross term and actually causes a precession of the node $\Omega$, which must be included in the total orbit element $\varpi$ (only the sum $\varpi = \omega + \Omega$ is relevant for coplanar orbits).  This term has the same origin as the de Sitter precession of the node of the Earth-Moon system induced by the Sun, which has been measured using lunar laser ranging.   

Since we worked to linear order in the perturbations due to the third body,  we can simply sum over all the other planets, to obtain
\begin{align}
\Delta \varpi &= \frac{6\pi Gm}{c^2 p} +  \frac{3\pi}{2} \sum_A \frac{m_A}{m} \left ( \frac{a}{R_A} \right )^3 (1-e^2)^{1/2} 
\nonumber \\
& \quad
\times \left [1 + \frac{1}{2} \frac{Gm}{c^2 a} \frac{28 + 47e^2}{(1-e^2)^{2}}\right ] 
\nonumber \\
& \quad
+ 4\pi \sum_A \frac{Gm_A}{c^2 a} \left ( \frac{a}{R_A} \right )^{5/2}\,. 
\label{eq1:delom2}
\end{align}
Inserting the relevant values for Mercury and the other planets out to Saturn, we obtain $42.98$ as/cy for the GR term, $384$ as/cy for the planetary perturbation coefficient and $4.2 \times 10^{-7}$ for the correction term inside the square brackets.  Thus the contribution of the cross-term perturbation at quadrupole order is $1.6 \times 10^{-4}$ as/cy, or $3.7 \times 10^{-6}$ of the main GR precession.  Notice that the planetary coefficient is smaller than the full planetary effect of $530$ as/cy because our quadrupole approximation underestimates the contributions from Venus and Earth.  The gravitomagnetic or de Sitter term contributes $6.4 \times 10^{-5}$ as/cy or $1.5 \times 10^{-6}$ of the main GR precession.

In principle, the calculations described here could be carried to higher order in the expansion of the perturbing fields of the planets.

\begin{table}[t] 
\centering
\caption{Contributions to Mercury's perihelion advance.} 
\vskip 12pt
\begin{tabular}{@{}lcc@{}} 
\hline 
Effect & Formula & Value relative\\ 
&(rad/orbit)&to GR\\
\noalign{\smallskip}
\hline 
\noalign{\smallskip}
Solar oblateness&$3\pi J_2 (R/p)^3$&$6.5 \times 10^{-4}$\\
\noalign{\smallskip}
Frame dragging&$-8\pi GJ/c^2(Gmp^3)^{1/2}$&$4.7 \times 10^{-5}$\\
\noalign{\smallskip}
PN cross term&see Eq.\ (\ref{eq1:delom2})&$3.7 \times 10^{-6}$\\
\noalign{\smallskip}
GM/de Sitter&see Eq.\ (\ref{eq1:delom2})&$1.5 \times 10^{-6}$\\
\noalign{\smallskip}
2PN&$-6\pi (Gm/2c^2p)^2 (10-e^2)$&$6.6 \times 10^{-8}$\\ 
\noalign{\smallskip}
\hline
\end{tabular}
\label{tab:mercury}
\end{table}

Table \ref{tab:mercury} lists the important subdominant contributions to Mercury's perihelion advance.  The solar oblateness contribution assumes a value of $J_2$ given by that inferred from helioseismology or from analyses of MESSENGER data.  The uncertainty in $J_2$ is around 10 percent.  The contribution of the dragging of inertial frames induced by the solar angular momentum $J$ is at the parts in $10^5$ level, while the leading cross-term effects are at parts in $10^6$.  The second post-Newtonian (2PN) contribution is significantly smaller, at parts in $10^8$.

\section{Calculations}

We begin with the PN $N$-body equations of motion in general relativity, as displayed in Eq.\ (9.127) of \cite{2014grav.book.....P}, Eq.\ (6.78) of \cite{tegp} or Eq.\ (6.79) of \cite{tegp2} (with PPN parameters chosen to be those of GR), truncated to three bodies.   We further restrict to a heirarchical triple system consisting of an inner binary with separation vector $\bm{x}_{12}$ and a distant third body of mass $m_3$ at a distance $R \gg r$.   The inner binary consists of a test mass (body 1) orbiting a central object (body 2) of mass $m$.   Thus the outer body's orbit is unaffected by the test body, and we choose that orbit to be circular and coplanar with the inner orbit.   We expand the vector $\bm{x}_{13}$  that joins the third body to the test mass in powers of $r/R$, with $r=|\bm{x}_{12}|$, $\bm{X} = \bm{x}_{23}$ and $R=|\bm{X}|$, retaining terms of order $Gm_3 r/R^3$ in the Newtonian acceleration, corresponding to quadrupole order, and keeping terms that scale as $[G^2 m_3 m /c^2 R^3] (r/R)^n$ in the PN accelerations for any $n \le 0$.  We exclude PN cross terms with $n >0$, as these will be progressively smaller than the terms being kept.   We also keep the conventional PN terms generated by the central mass, which scale as $G^2 m^2/c^2 r^3$.   The resulting equation of motion has the form
\begin{align}
{\bm a} &= - \frac{Gm{\bm n}}{r^2} - \frac{Gm_3 \,r}{R^3} \left [ {\bm n} -3({\bm n} \cdot {\bm N}) {\bm N} \right ]
+ \frac{1}{c^2}[{\bm a}]_{\rm Binary} 
\nonumber \\
& \quad + \frac{1}{c^2}[{\bm a}]_{\rm Cross}
+ O\left (\frac{G^2 m m_3 r}{c^2 R^4}\right ) \,,
\label{acc13}
\end{align}
where $\bm{N} = \bm{X}/R$, $\bm{n} = \bm{x}_{12}/r$, and 
\begin{align}
[{\bm a}]_{\rm Binary} &= \frac{Gm{\bm n}}{r^2} \left (\frac{4Gm}{r} - v^2 \right ) +  \frac{4Gm\dot{r} {\bm v}}{r^2} \,,
\nonumber
 \\
\left [{\bm a}\right]_{\rm Cross} &=
2\frac{G^2 mm_3}{R^3}\left [ {\bm n} -6({\bm n} \cdot {\bm N}) {\bm N} +3
 {\bm n} ({\bm n} \cdot {\bm N})^2 \right ]
 \nonumber \\
&
+ \frac{Gm_3 r}{R^3} \left [ 4{\bm v} \left \{ \dot{r} -3({\bm n} \cdot {\bm N})({\bm v} \cdot {\bm N}\right \} 
\right .
 \nonumber \\
& \qquad
\left .
- v^2 \left \{ {\bm n} -3({\bm n} \cdot {\bm N}) {\bm N} \right \} \right ]
 \nonumber \\
&
- \frac{Gm_3}{R^2} \left [ 4 \bm{v} \times ( \bm{N} \times \bm{V}_3 ) - 3 \bm{v} ( \bm{N} \cdot \bm{V}_3 ) \right ]
  \,,
\label{aPN3body}
\end{align}
where $\dot{r} = {\bm n} \cdot {\bm v}$.  There are additional PN cross terms that scale with the values $n = -5/2,\, -2$ and $-1$; these turn out to have no secular effect on the orbit elements of the inner binary, so we do not display them (see Eq.\ (4.7b) of \cite{2014PhRvD..89d4043W} for the full set of terms).

We now treat all but the Newtonian two-body acceleration as perturbations, and define the {\em osculating} orbit of the inner binary by (see Secs.\ 3.2 and 3.3 of \cite{2014grav.book.....P} for details)
\begin{align}
r & \equiv p (1+e \cos f)^{-1} \,,
\nonumber \\
\bm{v} &\equiv \dot{r} \bm{n} + \frac{\sqrt{Gmp}}{r} \bm{\lambda} \,,
\nonumber \\
\bm{n} &\equiv \cos (\varpi + f) \bm{e}_X + \sin (\varpi + f) \bm{e}_Y \,,
\nonumber \\
\bm{\lambda} &\equiv \partial \bm{n} / \partial f \,,
\nonumber \\
\frac{df}{dt} &\equiv \frac{\sqrt{Gmp}}{r^2} - \frac{d\varpi}{dt} \,,
\label{eq2:orbits}
\end{align}
where $f$ is the true anomaly and $\bm{e}_X$ and $\bm{e}_Y$ are fixed reference directions.  We then find the components of the perturbing accelerations along the $\bm{n}$, $\bm{\lambda}$ and $\bm{\hat h} = \bm{n} \times \bm{\lambda}$ directions, and insert them into the Lagrange planetary equations \cite{2014PhRvD..89d4043W}, which give equations for the evolution of the orbit elements $X^\alpha$ of the general form
\begin{equation}
\frac{dX^\alpha}{dt} = Q^\alpha (X^\beta(t), t) \,.
\end{equation}
We then integrate these equations to obtain secular variations of the orbital elements.  

However, in order to find the secular changes in the orbit elements induced by the cross terms in the equations of motion, we must carefully incorporate higher-order effects in the perturbation equations themselves.
First, the orbit elements $p$, $e$ and $\varpi$ vary periodically during the orbit.  Thus the PN-induced variations in these elements must be inserted back into the Newtonian perturbation terms generated by the third body, and the third-body induced variations must be inserted back into the perturbation terms generated by PN effects.  These will produce cross-term contributions of the same order as those coming directly from the equations of motion.
Second, it is conventional to identify secular variations by integrating over a complete cycle of the true anomaly, which runs from pericenter to pericenter.  But in converting from $d/dt$ in the Lagrange planetary equations to $d/df$,  we must use the last of Eqs.\ (\ref{eq2:orbits}) instead of the conventional relation $r^2 df/dt = (Gmp)^{1/2}$.  The added term comes from the fact that, while $t$ is measured from a fixed moment of time, $f$ is measured from the pericenter, which changes via $\dot{\varpi}$.    This added term, interacting with the PN and Newtonian third-body terms, will also generate cross-term effects between PN and third-body terms.   Finally, it is important to define consistently the orbit-averaged elements and the ``average-free'' variations of the elements over an orbital timescale; this is best carried out using a standard ``two-timescale'' analysis (see \cite{2004PhRvD..69j4021M} and \cite{2017PhRvD..95f4003W} for examples in a post-Newtonian context).   It is also conventional, in considering secular perturbations in a many-body context, to average over the orbital period of the third body.  The result of such an analysis is that, over one orbit, $\Delta p = \Delta e = 0$, and $\Delta \varpi$ is given by Eq.\ (\ref{eq1:delom1}).  

\section{Discussion}

The PN cross-term effects on Mercury's perihelion advance that we have pointed out arise from a subset of the post-Newtonian terms in the $N$-body equations of motion.  Those equations (modified to include the PPN parameters $\beta$ and $\gamma$) as adopted by Moyer in the early Jet Propulsion Laboratory (JPL) technical memoranda \cite{1971mfdo.book.....M,moyer2000} are the basis for many modern ephemeris and orbit-determination codes (see eg. \cite{2014IPNPR.196C...1F}).  However different groups or space agencies adopt different implementations of the basic equations.  If all ephemeris codes currently in use retain the summations over all the planets in all post-Newtonian terms, then, by definition all the relevant cross-term effects will be included, along with many effects that are negligible (such as PN effects due to the planets alone, of order $G^2 m_3^2 r^2/c^2 R^4$).   If there are any truncations of the sums, then the code might not properly account for the cross-term effects pointed out here.  The codes currently in use at JPL do include all terms \cite{Folkner}, but it is not known if this is universally true; it would be important to verify this, particularly for groups that will be involved in BepiColombo data analysis.   
Even if all such terms are included in the codes, their existence and cross-correlations may play a role in assessing the uncertainties in estimating $\gamma$ and $\beta$, and in measurements of the contributions to Mercury's perihelion advance arising from the solar quadrupole moment and from frame dragging that will be carried out using data from BepiColombo.   

We assumed general relativity in deriving the cross terms reported here; it is straightforward to generalize those results to the PPN formalism (eg with $\gamma$, $\beta$, $\alpha_1$ and $\alpha_2$ arbitrary), and those results will be reported elsewhere.  But the present constraints on these parameters are already so stringent that we do not expect PPN cross-term effects to contribute directly to improving the bounds on the PPN parameters.

Finally, at a purely pedagogical level, it is often stated that the relativistic perihelion advance of Mercury is really only a test of the vacuum  Schwarzschild solution (or of the slow rotation limit of the vacuum Kerr solution, if one wishes to include the frame-dragging effect), since all the relativistic effects can be derived simply from those metrics.   If BebiColombo can reach a part per million accuracy in measuring the perihelion advance, it will be possible to put this idea to rest, since it will measure, for the first time, {\em relativistic} effects on Mercury's orbit arising from the planets that surround it.

\acknowledgments

This work was supported in part by the National Science Foundation,
Grant No.\ PHY 16-00188.  We are grateful for the hospitality of  the Institut d'Astrophysique de Paris, where parts of this work were carried out.

%\bibliography{refsTEGP2}

\end{document}